\documentclass[dvips]{article}
\usepackage{icrctc07}

\title{OG 2.1-2.4: Gamma-ray Astronomy}

\authors{Jim Hinton}
\afiliations{
School of Physics \& Astronomy, University of Leeds, Leeds LS2 9JT, UK\\
}

\email{j.a.hinton@leeds.ac.uk}

\abstract{

The relevance of gamma-ray astronomy to the search for the origin of
the galactic and, to a lesser extent, the ultra-high-energy cosmic
rays has long been recognised. The current renaissance in the TeV
gamma-ray field has resulted in a wealth of new data on galactic and
extragalactic particle accelerators, and almost all the new results in
this field were presented at the recent International Cosmic Ray
Conference (ICRC).  Here I summarise the 175 papers submitted on the
topic of $\gamma$-ray astronomy to the 30$^{\rm th}$ ICRC in Merida,
Mexico in July 2007.

}

\begin{document}

\maketitle

\section{Introduction}

This paper reports on the results from the sessions OG 2.1--2.4 of the
30$^{\rm th}$ ICRC. These sessions covered topics related to the
origin of cosmic rays (CRs) as probed by $\gamma$-ray and X-ray
measurements.  In fact very few papers concerned purely with X-ray
measurements were presented and so for simplicity I will discuss only
the results involving $\gamma$-rays here. The classifications are
defined as follows:

\begin{itemize}
\item \emph{OG.2.1} Diffuse X-ray and gamma-ray emission
\item \emph{OG.2.2} Galactic sources (Binaries, pulsars, SN remnants, etc.)
\item \emph{OG.2.3} Extra-galactic sources (AGNs, Quasars, Gal.clusters, etc.)
\item \emph{OG.2.4} Gamma-ray bursts 
\end{itemize}

A total of 175 papers (including presentations and posters) where submitted
under these four sections, the vast majority (144) under \emph{OG.2.2} and 
\emph{OG.2.3}. There was also a predominance of contributions from experimental
collaborations involved with \emph{ground-based} $\gamma$-ray
astronomy (123/175).  I will therefore focus in this summary on
experimental results in TeV $\gamma$-ray astronomy. Indeed,
essentially all recent results in the $\gamma$-ray field were
presented at this conference. This is natural if one follows the
broadest possible definition of cosmic rays as simply ``astrophysical
relativistic particles'': $\ge$ GeV $\gamma$-rays can \emph{only} be
produced by CRs. Conversely, it is increasing recognised that
$\gamma$-ray measurements provide a powerful tool for studying the
acceleration and propagation of CRs of all energies.

After a brief summary of the instrumentation available for
$\gamma$-ray astronomy I will present my personal selection of
highlights in each of the sections listed above. I apologise in
advance to everyone whose work I have unfairly omitted and to all
whose work I may inadvertently misrepresent.


\section{Experimental Status}

Although the sessions \emph{OG 2.1 - 2.4} effectively cover only the
results from $\gamma$-ray detectors and not the status of these
instruments, it is useful to begin with a summary of the existing
instrumentation and the advantages and short-comings of different
approaches. For the moment, there is a clear division in the field
between measurements in the roughly 0.1--10 GeV range (High energy or
\emph{GeV} measurements) made with satellite based instrumentation and
roughly 0.1--100 TeV (very high energy, VHE, or \emph{TeV})
measurements made with ground-based instruments. A real overlap
between these domains will very likely be established within the next
few years, but for now they can be considered separately:

\subsection{GeV}

After a period of relative quiet, the GeV field is now increasingly
active as a consequence of the planned launch of the GLAST satellite
early in 2008 and the recent launch of AGILE. The upcoming new
instrumentation has prompted several authors to revisit the data of
the EGRET instrument (1991-2000).  Perhaps, the most significant of
these new analyses is the production of a new catalogue after modified
analysis and in particular modified diffuse background subtraction
\cite{155}. The new 3GR catalogue contains 23 new sources, but 121
third EGRET catalogue sources are not found in the new
analysis. Whilst this new analysis is controversial, there are
certainly indications that diffuse $\gamma$-ray background
uncertainties are such that the positions and even existence of many
3EG sources are very uncertain.  Whilst the better angular resolution
and sensitivity of GLAST with respect to EGRET will certainly help
with source identification, it is clear that understanding the diffuse
background is crucial to the success of GLAST for galactic
astrophysics. A major effort is underway in the GLAST collaboration to
improve models for the diffuse emission based on CR transport and
tracers for atomic and molecular target material and radiation fields
\cite{762,766}.  Amongst the presentations on the scientific potential
of GLAST were reviews of expectations for blazar measurements \cite{1211} 
and for detections of pulsars \cite{1286}, pulsar wind
nebulae and supernova remnants \cite{390}, 
and for GRBs \cite{1020}
and also possibilities for more exotic phenomena such as inverse
Compton halos around massive stars \cite{606}.



The relationship between sources at GeV and TeV energies was discussed
by several authors. A systematic comparison based on the region of the
H.E.S.S. galactic plane scan showed essentially no evidence for
correlation between the H.E.S.S. and 3EG catalogues \cite{391,392}.
The lack of sensitivity of EGRET seems to be a major factor in the
non-detection of TeV sources and GeV energies, whereas the existence
of spectral breaks (or cut-offs) is likely required to explain the
missing GeV sources at TeV energies. Another complicating factor in
the comparison of GeV and TeV data is the mismatch in field-of-view
and angular resolution for existing measurements.  The possibility
that some 3EG sources are perhaps rather extended and hence difficult
to detect with narrow field-of-view Cherenkov telescopes was raised by
the MILAGRO collaboration (MILAGRO has a very wide field of view and
modest angular resolution). Indeed, there are hints of a connection
between the new MILAGRO sources and 3EG/GeV sources \cite{735}, but
without better angular resolution measurements, the problems of source
identification will likely remain.

The AGILE satellite, a relatively small area, but wide field of view
instrument, was launched in April 2007. The energy range and overall
sensitivity of AGILE are comparable to EGRET, but its wider field of
view makes it particularly suitable for the monitoring of blazars, and
it could prove useful as a trigger for TeV instruments \cite{363}.

\subsection{TeV}

\begin{figure}[t]
\noindent
\hspace{-12mm}\includegraphics[width=0.45\textwidth, angle=270]{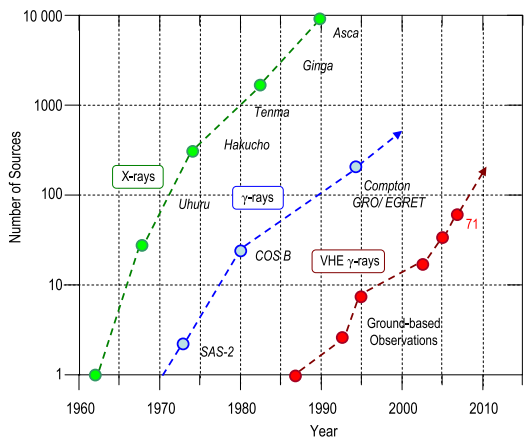}
\caption{Source numbers versus time in the X-ray, high-energy $\gamma$-ray
and VHE $\gamma$-ray domains (adapted from a plot by Tadashi Kifune).
VHE source counts plotted are those reported by rapporteurs at each international
cosmic ray conference.
}\label{fig:kifune}
\end{figure}

Ground-based techniques for $\gamma$-ray astronomy rely on the
development of cascades (air-showers) initiated by astrophysical
$\gamma$-rays. Such cascades only persist to ground-level above
$\sim$1 TeV and only produce significant Cherenkov light above a few
GeV, setting a fundamental threshold to the range of this
technique. Ground-based measurements in the $\sim$50~GeV to $\sim$100
TeV range have resulted in a exponential increase in the number of
sources known to emit in this energy range over the last few
years. This progress is compared to that in other energy ranges in
figure~\ref{fig:kifune}, an updated version of a plot due to Tadashi
Kifune. Two principal approaches to such measurements exist and this
are considered here in turn.


\subsubsection{Cherenkov Telescopes}

The most successful approach to ground-based $\gamma$-ray 
astronomy is that based on the imaging of the Cherenkov light
produced by photon initiated cascades in the Earth's atmosphere.
These relatively small field-of-view instruments ($\sim$4$^{\circ}$) 
have $\sim$10\% duty cycle due the need for good weather and 
complete darkness, but achieve angular and energy resolution
much better than that of any other ground-based $\gamma$-ray technique 
($\sim$0.1$^{\circ}$ and $\sim$15\% respectively).
The success of this technique also results from the ability to 
reject a large fraction of the cosmic-ray background based on
the shape of the Cherenkov images (see for example \cite{hillas}).
The use of multiple Imaging Atmospheric Cherenkov Telescopes (IACTs) 
to allow stereoscopic reconstruction of the 
shower provided a further breakthrough in sensitivity and resolution.
The existing Cherenkov telescopes and telescope arrays are
summarised in table~\ref{tab:inst}. Three multiple-telescope arrays
of IACTs are currently operating:
VERITAS \cite{810}, 
CANGAROO-III \cite{166} and H.E.S.S.

H.E.S.S. is a four telescope array located in the Khomas highlands of
Namibia. The latitude of H.E.S.S., its relatively wide field of view
($5^{\circ}$) and its unprecedented sensitivity (0.7\% of the flux
from the Crab Nebula at $5\sigma$ in 50 hours of observations) make it
an ideal instrument to survey the galactic plane. Indeed, the ongoing
H.E.S.S. galactic plane survey has led to a dramatic increase in the
number of galactic TeV sources \cite{269}.

The recently completed (April 2007) VERITAS array is rather similar 
to H.E.S.S. in several respects and can be considered as a 
complementary northern hemisphere instrument. Despite its 
recent completion VERITAS has already produced several important results
(summarised in \cite{810}).

CANGAROO-III consists of three new telescopes deployed around the
single CANGAROO-II telescope in Australia. Some controversy surrounded
certain sources detected using CANGAROO-I and -II and subsequent
non-detections using H.E.S.S. As of this conference none of these
disagreements remain following more sensitive observations with
CANGAROO-III and resulting retraction of some earlier results
\cite{166}. In addition CANGAROO-III has been used to confirm some
of the discoveries using H.E.S.S. \cite{477,320}.

\begin{table*}
\begin{center}
\begin{tabular}{|l|c|c|c|c|c|c|c|c|c|c|} \hline
Instrument & Lat. & Long. & Alt. & Tels. & Tel. Area & Total A. & Pixels & FoV & Thresh. \\
           & ($^{\circ}$) & ($^{\circ}$) & (m) &            & (m$^{2})$  & (m$^{2}$)  &    & ($^{\circ}$) & (TeV)  \\\hline 
H.E.S.S. & -23 & 16 & 1800 & 4 & 107 & 428 & 960 & 5 & 0.1  \\ 
VERITAS & 32 & -111 & 1275 & 4 & 106 & 424 & 499 & 3.5 & 0.1  \\ 
MAGIC & 29 & 18 & 2225 & 1 & 234 & 234 & 574 &  3.5$^{\dagger}$ & 0.06 \\ 
CANGAROO-III & -31 & 137 & 160 & 3 & 57.3 & 172 & 427 & 4 & 0.3 \\
Whipple & 32 & -111 & 2300 & 1 & 75 & 75 & 379 & 2.3 & 0.3 \\
Shalon & 43 & 77 & 3338 & 1 & 11.2 & 11.2 & 144 & 8 & 0.8 \\
TACTIC & 25 & 78 & 1300 & 1 & 9.5 & 9.5 & 349 & 3.4 & 1.2 \\\hline
\emph{HEGRA} & 29 & 18 & 2200 & 5 & 8.5 & 43 & 271 & 4.3 & 0.5 \\
\emph{CAT} & 42 & 2 & 1650 & 1 & 17.8 & 17.8 & 600 & 4.8$^{\dagger}$ & 0.25 \\
\hline
\end{tabular}
\end{center}
\caption{Principle characteristics of currently operating (and
selected historical) IACTs and IACT arrays. The
energy threshold given is the approximate trigger-level (rather than
post-analysis) threshold for observations close to zenith. 
$^{\dagger}$ These instruments have pixels of two different
sizes.  }
\label{tab:inst}
\end{table*}

The 17~m diameter MAGIC telescope on La Palma represents the
state-of-the-art in terms of single dish instruments. The instrument
is optimised for low energy measurements and has had considerable
recent success in discovering steep spectrum extragalactic sources.

Following the success of H.E.S.S. and MAGIC, both instruments are
in a second phase of construction. For H.E.S.S. this involves the
construction of a 600 m$^{2}$ telescope at the centre of the existing
array, with the aim of achieving useful sensitivity in the unexplored
$<$50~GeV region. MAGIC phase-2 consists of the construction of a
second 17~m diameter telescope with the aim of using stereoscopic
techniques to improve sensitivity and reduce threshold.

Useful contributions are also being made using instruments of more
modest sensitivity such as TACTIC \cite{170} and the long serving
Whipple 10~m telescope \cite{728}.  Both these instruments are being
used to monitor the brightest TeV blazars and can be used to alert
more sensitive instruments.


In addition to these imaging telescopes, several groups have made use
of non-imaging Cherenkov telescopes, these include the PACT array and
several groups making use of modified solar power facilities.  The
enormous available mirror area of these facilities can be used at
night to conduct air-Cherenkov based $\gamma$-ray measurements. The
CELESTE, STACEE, Solar-2, CACTUS and GRAAL collaborations all pursued
this concept and were largely successful in achieving low ($<$100~GeV)
energy thresholds but unfortunately their discovery potential was
limited by relatively poor background rejection capabilities (in
comparison to imaging techniques). To my knowledge none of these
instruments is still operational.  The final results from the recently
decommissioned STACEE instrument were presented
here. 

\subsubsection{Shower Particle Detectors}

The depth of maximum development of photon initiated air-showers
typically occurs close to 10~km a.s.l. for 1~TeV
$\gamma$-rays. However, the tail of the shower is detectable far past
maximum for detectors with sufficient collection area. These
instruments achieve duty-cycles close to 100\% and $\sim$1~sr
field-of-view (FoV), but have modest angular and energy resolution
($\sim$1$^{\circ}$ and $\sim$50\% respectively).  Two approaches exist
for $\gamma$-ray measurements via direct sampling of the shower
particles.  The classical method is to use an array of (relatively)
widely spaced scintillator-based detectors. The Tibet AS$\gamma$
instrument employs this approach at high altitude (4300~m) to reduce
the energy threshold to $\sim$3~TeV. The second approach requires
complete coverage of the ground to ensure efficient collection of
shower particles and hence lower energy threshold. The recently
completed ARGO-YBJ detector \cite{1029} at the Tibet site is a
solid-state detector following this approach. Arguably the most
successful shower-particle detector built for $\gamma$-ray astronomy
is MILAGRO, a water-Cherenkov based instrument at Los Alamos (2630 m
altitude). This instrument has been operating for 7 years, but recent
detector and analysis improvements have led to the significant
detection of 4 sources including 3 new discoveries \cite{735}. The new
analysis cuts significantly improve background rejection power, but at
the expense of increased energy threshold (to $\sim$20~TeV from a
trigger threshold of $\sim$1~TeV).  The MILAGRO instrument is nearing
the end of its operational life, but plans for a follow-up
instrument built at much higher altitude are well advanced
\cite{1238,1250}. The High Altitude Water Cherenkov (HAWC) instrument
should reach significantly lower energies and better sensitivity
whilst maintaining the advantages of high duty cycle and FoV.

During the 1990s several more widely spaced ground arrays were
constructed to search for $\sim$100~TeV $\gamma$-rays.  The very long
exposures of these instruments partially compensates for the low
$\gamma$-ray rates at these energies and the absence of significant
background rejection capabilities. The upper limits presented by the
CASA-MIA \cite{814} and SPASE-2 \cite{678} collaborations therefore
lie at interesting flux levels.  The GRAPES-III instrument is an
intermediate case with a $\sim10$ TeV threshold \cite{1054}.

\section{OG 2.1: Diffuse Gamma-ray Emission}

Particles (particularly protons and nuclei) of $\ge$ GeV energies, can
readily propagate very large distances in the ISM without significant
energy losses. As a consequence the emission associated with these
energy losses is often rather diffuse.  At GeV energies the
$\gamma$-ray sky is dominated by the diffuse emission produced by
galactic cosmic-rays in the ISM. At TeV energies it appears that the
flux of the diffuse component is comparable with that from discrete
sources \cite{kappes}. This is unsurprising as the typical energy
spectra of discrete sources lie close to the test-particle shock
acceleration spectrum of $E^{-2}$ (the mean photon index of sources
found in the H.E.S.S. galactic plane survey was 2.3
\cite{HESS:scanp2}) and the spectrum of high energy $\gamma$-rays
produced in hadronic interactions in the ISM approximately follows
that of the incident protons and nuclei i.e.$E^{-2.7}$. The lower
relative flux of the diffuse component and the small FoV of the most
sensitive TeV instruments, make measurements of the galactic TeV
diffuse emission very difficult.  The only existing measurement of the
(large-scale) diffuse TeV emission comes from the MILAGRO instrument
\cite{654}
\footnote{a localised measurement of diffuse emission has been 
made in the Galactic Centre, see below}.  The MILAGRO collaboration
have detected emission along the plane with localised enhancements
which have been identified as sources. After subtraction of these
sources the remaining emission roughly follows the distribution of
target material in the galaxy and is identified as diffuse
emission. The flux level of this emission lies about a factor two
above the predictions of the GALPROP model \cite{GALPROP} with
parameters tuned to best reproduce the data from EGRET.  As it seems
very likely that there is still a significant contribution from
discrete sources to this measurement, this level of agreement with
predictions seems satisfactory.

The MILAGRO collaboration also presented the results of a search for
intermediate scale ($>10$ deg) features over the whole sky
\cite{672}. Significant anisotropies are indeed seen, but appear
stronger in data without $\gamma$-ray selection cuts, suggesting they
are charged particle anisotropies, perhaps related to the tail-in
anisotropy seen using the Tibet AS$\gamma$ instrument
\cite{Tibet:TailIn} (and as such lie beyond the scope of this
summary).

\section{OG 2.2: Galactic Sources}

It is well established that the bulk of the cosmic rays 
measured at the Earth must originate within our own galaxy.
As a consequence those CRs with energies up to at 
least $10^{15}$ eV are often referred to as the \emph{galactic 
cosmic rays} (GCRs). The principal acceleration sites of the protons and 
nuclei of the GCRs are not yet well established. Indeed,
although they make up a small fraction of the total energy in
cosmic-rays, the origin of the electron component is equally
unclear and important to establish.

It has long been recognised, see for example \cite{Ginzburg},
that $\gamma$-ray measurements can aid in the identification 
of the CR acceleration sites in our galaxy. Two 
principal $\gamma$-ray production mechanisms are discussed here:
The decay of neutral pions produced in hadronic interactions,
which traces the product of ambient density and the density of
CR protons and nuclei, and Inverse Compton up-scattering
of ambient photon fields, tracing high energy electrons.

Although many TeV $\gamma$-ray sources are now known there are 
two major challenges to overcome to make progress in 
addressing the questions of cosmic-ray origin. The first and
most basic is to identify $\gamma$-ray sources with counterpart
objects at other wavelengths. This process can be far from 
straight-forward and many different techniques have been applied
to provide solid source identifications. Table~\ref{tab:gal} lists
the small fraction of galactic TeV sources with such identifications
(note that the selection is somewhat subjective and the list given
here is rather conservative). The second challenge is to infer
the nature, and spatial and energy distributions of, the primary
CRs. Differentiating between electrons and protons as
the radiating particles has proved difficult, although several
cases now exist where one or the other is strongly favoured.  


\begin{table*}
\begin{center}  
  \begin{tabular}{|l|l|l|l|l|l|l|l|} \hline 
Object	&  Discovered     &Year	&  Type	      & Method	& Flux  & Contrib.  \\\hline
PSR B1259$-$63  &  HESS	  & 2005 &  Binary     &Pos/Var & 7$^{\star}$	& \cite{553} \\    
LS\,5039	&  HESS	  & 2005 &  Binary     &Pos/Per	& 3$^{\star}$	& \cite{1305} \\
LS\,I\,+61 303  &  MAGIC  & 2006 &  Binary     &Pos/Var	& 16$^{\star}$	& \cite{700,351,574}\\
RX\,J1713.7$-$3946 & CANGAROO & 2000 &  SNR Shell  &	Mor	& 66	& \cite{524} \\
Vela Junior	&  CANGAROO & 2005&  SNR Shell  &	Mor	& 100	& \\
RCW\,86		&  HESS	  & 2007 &  SNR Shell  &	Mor  & $\sim$10 & \cite{280} \\
Cassiopeia A	&  HEGRA  & 2001 &  SNR	      &	Pos	& 3	& \cite{528}	  \\
Crab Nebula	&  Whipple& 1989 &  PWN	      &	Pos	& 100	& \cite{1078,986,954,1209}...\\
MSH\,15-52	&  HESS	  & 2005 &  PWN	      &	Mor	& 15	& \cite{422}  \\ 
Vela\,X		&  HESS	  & 2006 &  PWN	      &	Mor	& 75	&	  \\
HESS\,J1825$-$137 &  HESS	  & 2005 &  PWN	      & EDMor	& 12    & \cite{389}	  \\
PSR\,J1420$-$6049 &  HESS	  &2006&  PWN 	      &	Mor	&  7	&	  \\
The Rabbit	&  HESS	  &2006&  PWN	      &	Mor	&  6	&	  \\
G\,0.9+0.1	&  HESS	  &2005&  PWN	      &	Pos     &  2    &         \\
\hline
\end{tabular}
\end{center}
  \caption{ Galactic VHE $\gamma$-ray sources with well established
  multi-wavelength counterparts.  The instrument used to discover the
  VHE emission is given together with the year of discovery. Fluxes
  are approximate values expressed as a percentage of the flux from
  the Crab Nebula above 1~TeV, $^{\star}$ indicates variable emission.
  These associations were established through a range of methods,
  which are given in the table in abbreviated form: \emph{Pos}: The
  position of the centroid of the VHE emission can be established with
  sufficient precision that there is no ambiguity as to the low energy
  counterpart. In practise this is usually only possible for
  point-like sources. \emph{Mor}: There is a match between the
  $\gamma$-ray morphology and that seen at other (usually X-ray)
  wavelengths. This requires sources extended well beyond the typical
  angular resolution of IACTs ($\sim$0.1$^{\circ}$). \emph{EDMor}:
  Energy-dependent morphology which approaches the position/morphology
  seen at other wavelengths at some limit, and is consistent with our
  physical understanding of the source. \emph{Var}: $\gamma$-ray
  variability correlated with that in other wavebands. \emph{Per}:
  periodicity in the $\gamma$-ray emission matching that seen at other
  wavelengths. Note that all these objects have associated X-ray
  emission which has been interpreted as synchrotron radiation.
  Notable omissions from this table include Cyg\,X-1, IC~\,443 and
  W\,28. These objects are discussed in detail in the main text.
}
\label{tab:gal}
\end{table*}

\begin{figure*}[t]
\begin{center}
\noindent
\includegraphics[width=0.99\textwidth]{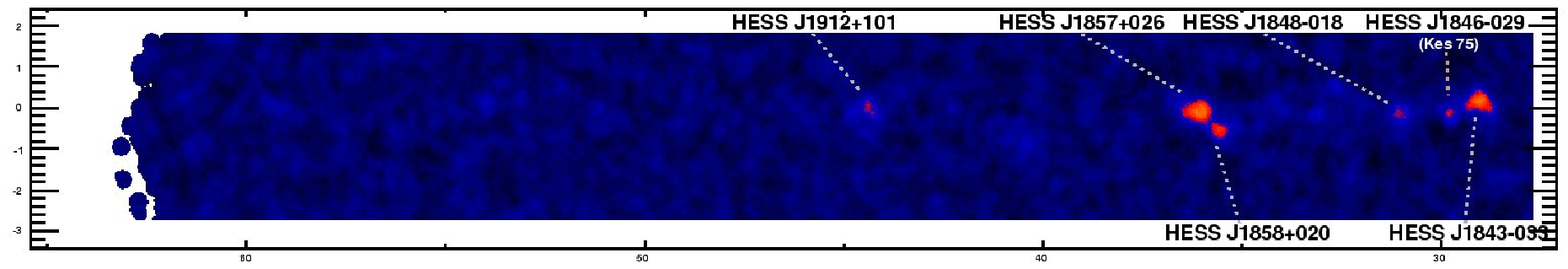}
\includegraphics[width=0.99\textwidth]{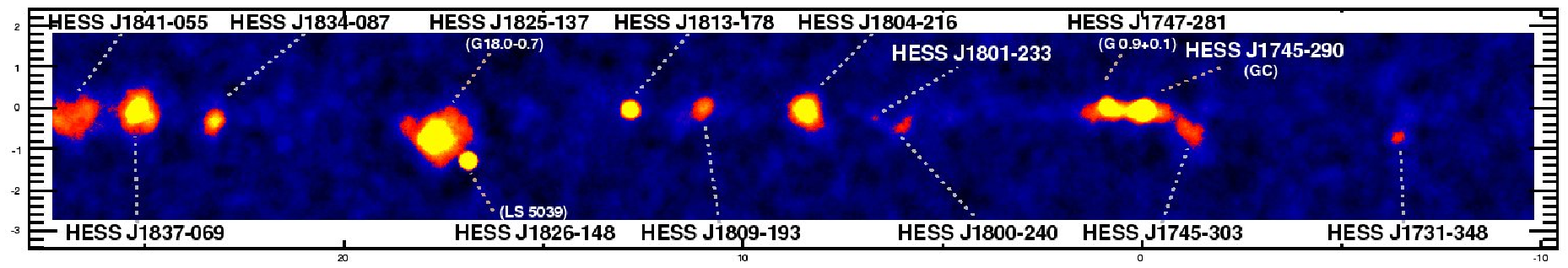}
\includegraphics[width=0.99\textwidth]{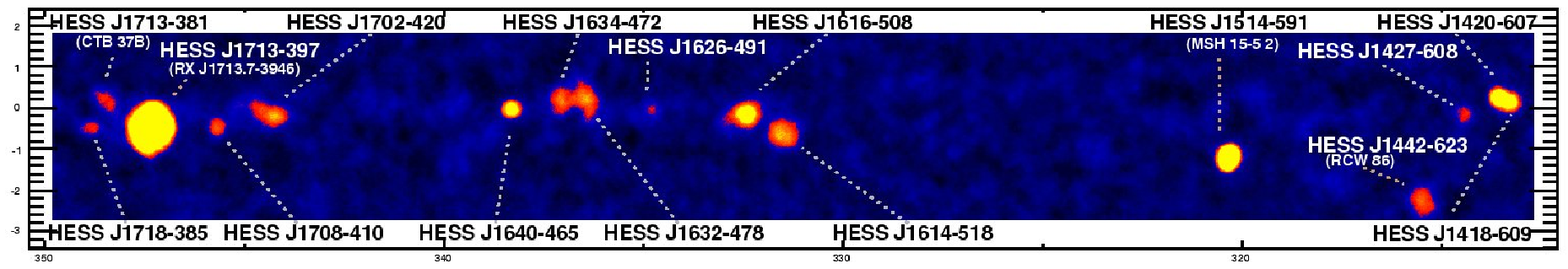}
\includegraphics[width=0.99\textwidth]{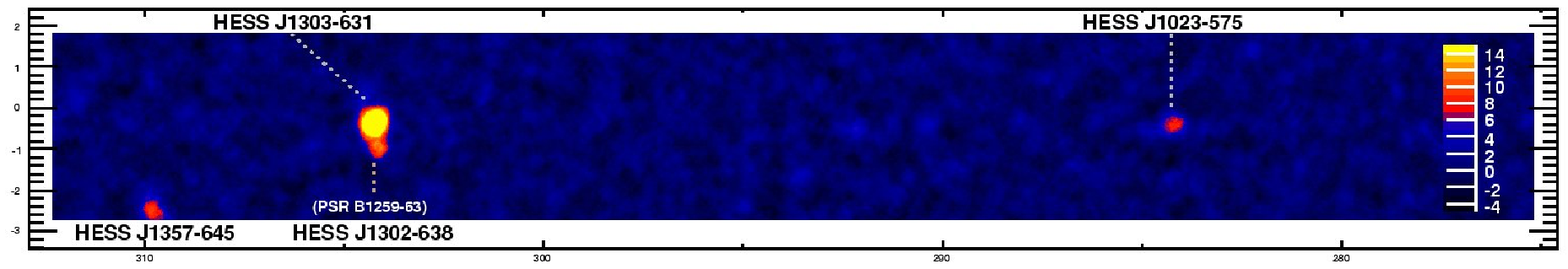}
\caption{The H.E.S.S. survey of the inner galaxy in $\sim$1~TeV $\gamma$-rays.
The colour-scale indicates the statistical significance for somewhat extended sources.
Image courtesy of the H.E.S.S. Collaboration.}\label{fig:scan}
\end{center}
\end{figure*}

Several of the sources in table~\ref{tab:gal} were discovered in the
survey of the galactic plane conducted by the H.E.S.S. collaboration
\cite{HESS:scanp2}.  The extension of this survey to cover essentially
the whole inner galaxy: $-85^{\circ} < l < 60^{\circ}, -2.5^{\circ} <
b < 2.5^{\circ}$ is responsible for many of the new sources summarised
here \cite{269} (see figure~\ref{fig:scan}).  The positive galactic
latitude extent of this survey is now limited by zenith angle
constraints.  The region inaccessible to H.E.S.S. has been covered by
MILAGRO measurements and a survey of the Cygnus region with VERITAS is
underway. It is to be hoped that by the time of the next ICRC a
complete sensitive survey of the galactic plane will exist, allowing
studies of the populations of galactic TeV sources.  The current
experimental situation already allows detailed studies of several
classes of galactic object, and these are considered here in turn.

\subsection{Supernova Remnants}

Supernova remnants (SNRs) have long been the prime candidates for the
acceleration of the bulk of the galactic cosmic ray protons and
nuclei.  They have sufficient energy, providing 10\% of the kinetic
energy of an average supernova explosion can be converted into
relativistic particles (see e.g. \cite{Ginzburg}), and there is a well
established mechanism: diffusive shock acceleration in the SNR shell
\cite{Bell:DSA,Blandford:DSA}.
Despite this, only rather recently has strong evidence for the
acceleration of particles in SNR shells begun to emerge. The
acceleration of $\sim$100~TeV electrons in SNRs was first suggested by
the interpretation of non-thermal X-ray emission from objects such as
SN\,1006 as synchrotron radiation \cite{ASCASN1006}.  The first
unambigous evidence for the existence of $>$ TeV particles in
supernova remnants come with the CANGAROO detection of
RX\,J1713.7$-$3946 \cite{CANGAROO:1713} and the subsequent higher
angular resolution measurements with H.E.S.S. which resolved the shell
in $\gamma$-rays ~\cite{HESS:1713}.  The current challenges in the
field are the expansion of the catalogue of TeV SNRs and the detailed
study of the brighest objects, to identify the nature of the radiating
particles (protons and nuclei or electrons).

\begin{figure*}[t]
\begin{center}
\noindent
\includegraphics [width=0.99\textwidth]{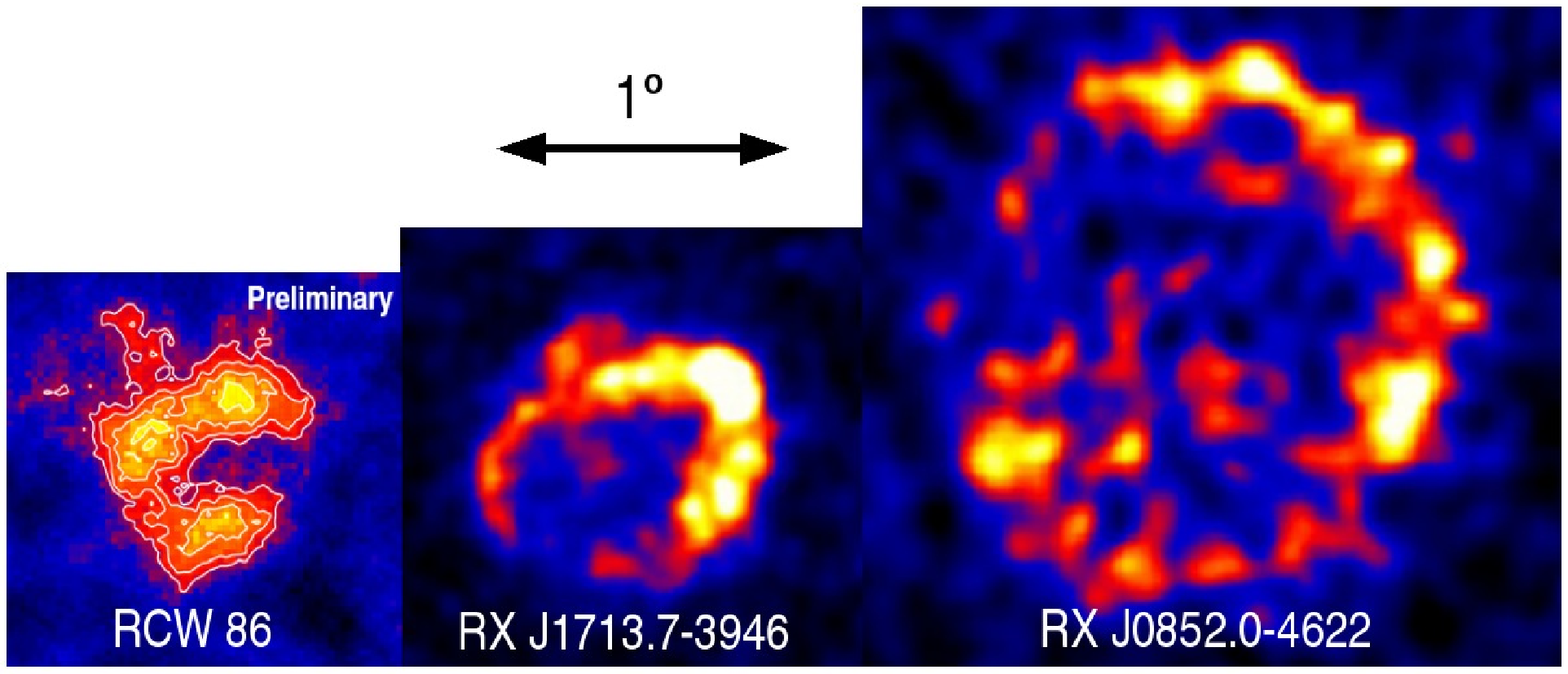}
\caption{The known shell-type $\gamma$-ray SNRs: RCW\,86 \cite{280}, 
RX\,J1713.7$-$3946 \cite{524}
and RX\,J0852.0$-$4622 (\emph{Vela Junior}) \cite{HESS:velajnr2}. 
All images are smoothed and were obtained using H.E.S.S.}\label{fig:3snr}
\end{center}
\end{figure*}

The progress in this area since the last ICRC has been considerable.
Three new TeV $\gamma$-ray sources associated with SNRs were presented
together with further data on all the known $\gamma$-ray emitting
SNRs.  Those VHE SNRs with apparent shell-type morphology are shown in
figure~\ref{fig:3snr}.  RCW\,86 is the weakest and most recently
discovered of these objects \cite{280}.  Recent X-ray measurements
suggest that RCW~86 is the remnant of the supernovae of 185~AD
\cite{vink}, placing it in the age range of the other TeV emitting SNRs.
The $9.4\sigma$ H.E.S.S. detection shows evidence for a shell roughly 
matching the X-ray morphology of this object. Unfortunately, due to its lower flux, 
it will be very difficult to study this object in the same level of detail
as RX\,J1713.7$-$3946 and RX\,J0852.0$-$4622. 

The two other newly discovered SNRs: IC\,443 and W\,28, both appear to
have emission correlated with available target material rather than
with the radio/X-ray emission of the SNR shell itself, suggesting that
the TeV emission may arise from interactions of hadronic CRs in (and
surrounding) the SNRs.  Both are also somewhat older than the
shell-type TeV SNR of figure~\ref{fig:3snr} (W\,28: $\sim$10$^{5}$
years, IC\,443: $\sim$3$\times10^{4}$ years).  The H.E.S.S. data on
W\,28 indicate at least 3 separate peaks in the emission, one
coincident with the brightest part of the radio shell (and the EGRET
source 3EG\,J1800$-$2338), but with the others lying outside the
shell, in coincidence with molecular clouds seen in $^{12}$CO data
\cite{129}.  TeV emission coincident with IC\,443 was recently
discovered independently by both the MAGIC \cite{487} and VERITAS
\cite{1170} collaborations. The $\gamma$-ray signal is at a
significance of $5.7\sigma$ in 29~hours of MAGIC data, and $7.1\sigma$
in 16~hours of VERITAS observations. The centroid of the emission is
consistent between the two measurements and is not coincident with the
X-ray PWN within the remnant, nor with the SNR shell, but rather with
a dense region towards the centre of the remnant (in
projection). Maser emission tracing dense shocked gas is coincident
with the emission, providing strong evidence that the signal arises in
the interaction of CRs accelerated in the shell interacting with
molecular material.  There is no evidence so far for spatial extension
of the signal, providing a motivation for deeper observations as
morphology matching that of the molecular clouds would confirm this
interpretation.

\begin{figure}
\begin{center}
\includegraphics[width=0.49\textwidth]{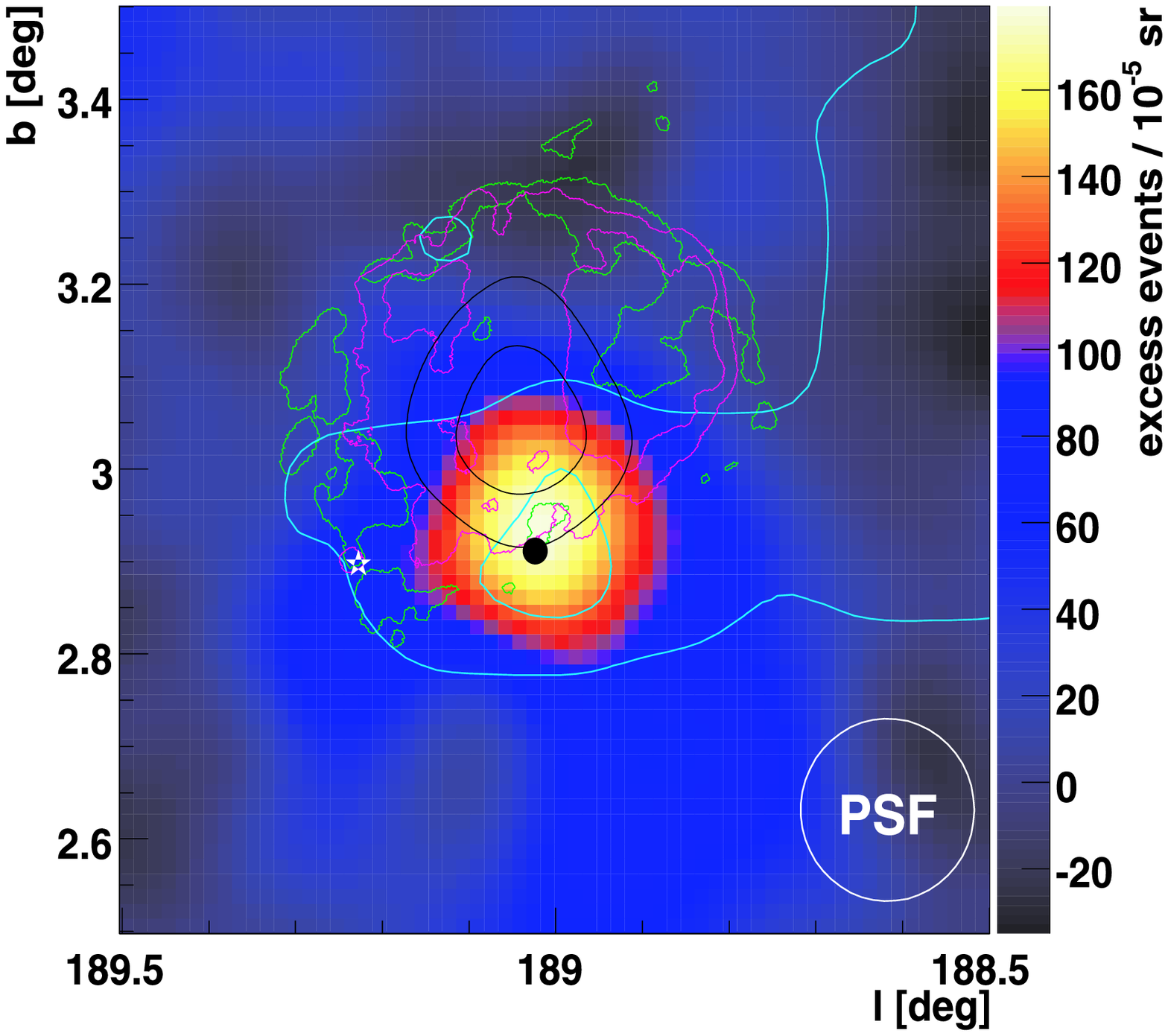}
\caption{$\gamma$-ray image of IC\,443 as seen by MAGIC above 150~GeV (colour scale, reproduced from \cite{487}).
Overlayed contours show: $^{12}$CO emission (cyan), 20 cm VLA data
(green), X-ray emission as seen using ROSAT (purple) and confidence
contours for the position of the EGRET source 3EG\,J0617+2238 (black).
The star shows the position of the PWN CXOU\,J061705.3+222127. The
black dot marks the position of a 1720 MHz OH maser. See \cite{487}
for details and references.  }
\label{fig:ic443}
\end{center}
\end{figure}

The two established TeV SNRs for which new $\gamma$-ray data were
presented are RX\,J1713.7$-$3946 and Cassiopeia A. Three years of
H.E.S.S. observations of the $\gamma$-ray bright SNR
RX\,J1713.7$-$3946 have resulted in spectral and morphological data
with very small statistical errors \cite{524}.  The energy spectrum of
RX\,J1713.7$-$3946 now spans from 0.3 to 80 TeV with a very
significant ($4.8\sigma$) signal above 30~TeV. This wide spectral
coverage provides a much greater challenge to modellers then previous
spectra, it now seems that inverse Compton scenarios for the emission
are becoming unlikely, whilst a hadronic origin of the emission is
favoured.  The young and radio-bright SNR Cassiopeia A was first
detected at TeV energies using the HEGRA telescope
array~\cite{HEGRA:CasA} at the 5$\sigma$ level in 232 hours of data
spread over several years of observations. This signal has now been
confirmed using the MAGIC telescope \cite{528}, at the $5.2\,\sigma$
level using 47 hours of observations. The MAGIC photon index of
$2.4\pm0.2$ is consistent with that measured using HEGRA:
$2.5\pm0.4$. The radio size of Cas A ($4'$) means that VHE morphology
of this object cannot be resolved with current instruments, but
further spectral measurements with MAGIC (and VERITAS) may be very
important.

The theory of particle acceleration in supernova shocks has been under
continuous development for the last 30 years. The principal
theoretical contributions in this area to this conference were those
of Berezhko, V\"olk and Ksenofontov on the SNRs: Tycho \cite{127},
Kepler \cite{126}, 
SN\,1987A \cite{125}, 
RX\,J1713.7$-$3946 \cite{614}
and Vela Junior \cite{597}. In the cases where upper limits to the TeV
emission exist (such as for Kepler's SNR) consistency with the
non-linear model can be used to provide density and distance
constraints. In those objects with measured TeV emission, the X-ray
and $\gamma$-ray data appear consistent with the picture of shocks
modified by hadronic CRs and $\gamma$-ray emission dominated by
neutral pion decay.  Further theoretical work involved more detailed
treatment of hadronic interactions and the inclusion of nuclei in the
calculation of $\gamma$-ray spectra \cite{681};  
the study of SNR evolution in a non-uniform medium \cite{1175};
and the possibility of `Jitter' rather than synchrotron X-ray 
emission dominating in SNR \cite{1268}.

\subsection{Pulsars and Pulsar Wind Nebulae}

\begin{figure}
\begin{center}
\includegraphics[width=0.49\textwidth]{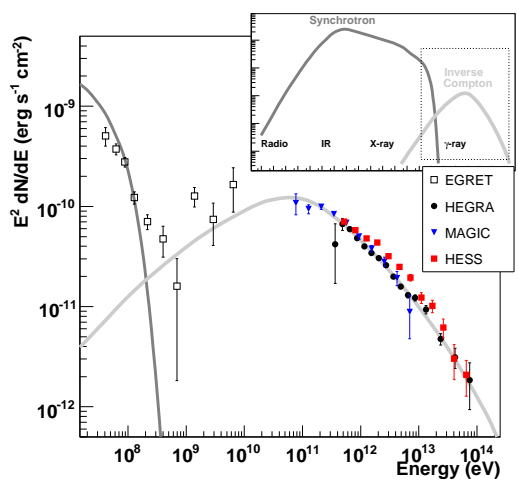}
\caption{The spectral energy distribution (SED) of the Crab Nebula. 
The inset shows a model radio-TeV SED with synchrotron and inverse Compton
components from \cite{HEGRA:crab}, the main panel shows the $\gamma$-ray part 
of the spectrum. EGRET and HEGRA data are reproduced from 
\cite{HEGRA:crab}, 
H.E.S.S. \cite{986} 
and MAGIC \cite{1078} data are those presented at this conference. 
}
\label{fig:crabspec}
\end{center}
\end{figure}

The Crab Nebula was the first TeV $\gamma$-ray source to be discovered
\cite{Whipple:crab} and is still the brightest steady and point-like
source in the TeV sky. The $\gamma$-ray emission from the Crab is
dominated by the pulsar below GeV energies and by steady emission from
the Nebula above. Figure~\ref{fig:crabspec} shows the broad-band
spectral energy distribution of the Crab Nebula, illustrating the
double-peaked emission common to all pulsar wind nebulae (PWN). The
two components are commonly attributed to synchrotron and
inverse-Compton scattering of a population of ultra-relativistic
electrons emerging from the termination shock of the pulsar wind.

The Crab Nebula is commonly used as a reference source in VHE
$\gamma$-rays and to verify the sensitivity of instruments as
predicted by Monte-Carlo simulations. Contributions from 
the VERITAS and MAGIC collaborations quote their sensitivities
as 31$\sigma/\sqrt{\rm hour}$ \cite{1290} (with 3/4 telescopes
operational) and 19$\sigma/\sqrt{\rm hour}$ \cite{1078}, respectively.
The newly commissioned  ARGO YBJ detector \cite{1029}, presented 
a $5 \sigma$ on the Crab Nebula in 50 days, which compares 
favourably to the roughly $\sim$2$\sigma/\sqrt{\rm 50\,days}$ signal
of MILAGRO (averaged over the full 7 year exposure).
Seven years of Crab data from the Whipple 10~m telescope were also presented,
illustrating the stability of this instrument~\cite{543}. 
Beyond its role as a calibration source, the Crab pulsar and its nebula 
are also of great interest astrophysically and several new spectral 
measurements of the Nebula were presented at this conference.
The new H.E.S.S. measurements~\cite{986} extend the spectrum up to
$\sim$80~TeV and at the low energy end, the MAGIC spectrum extends down to
$\sim$80~GeV \cite{1078} (see figure~\ref{fig:crabspec}).  
There is evidence for curvature in both data sets, with the MAGIC data 
being used to constrain the position of the high-energy peak 
in the spectral energy distribution to be 
$77\,\pm\,47$ GeV. The MILAGRO collaboration also presented a spectral
measurement for the Crab Nebula, the first measurement of its kind for
this instrument~\cite{710}.

\begin{figure*}
\begin{center}
\includegraphics[width=0.45\textwidth]{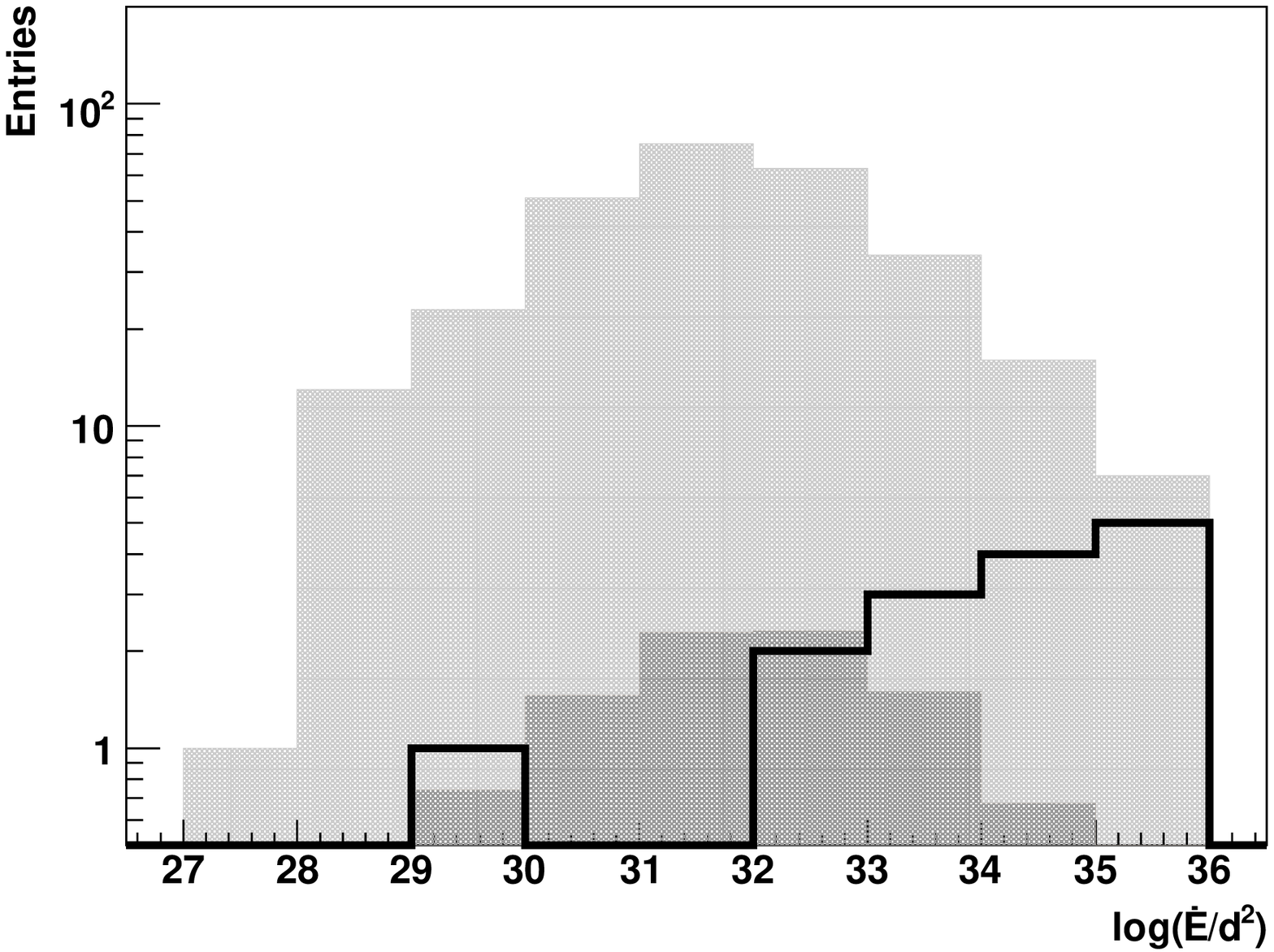}\includegraphics[width=0.45\textwidth]{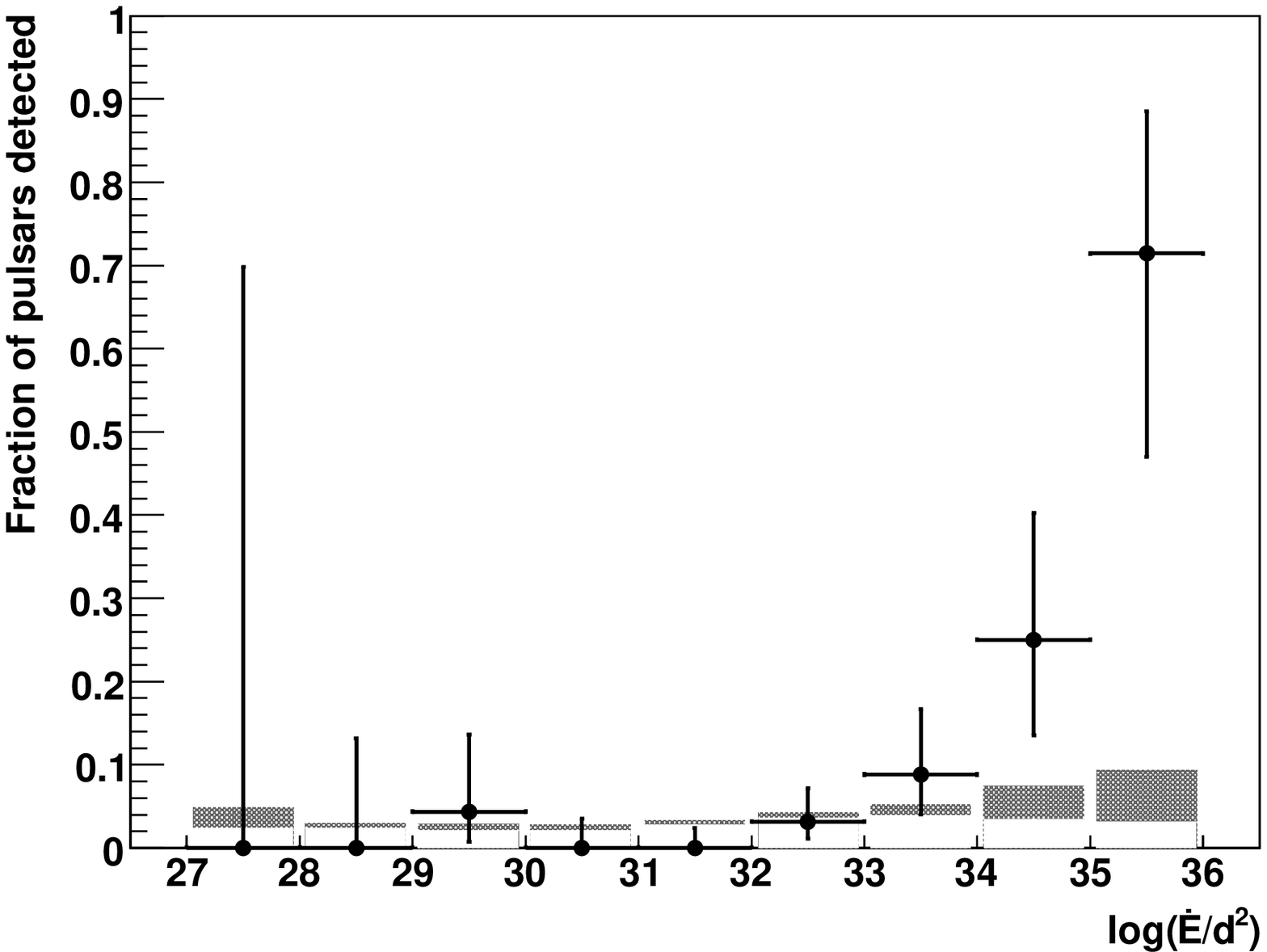}
\caption{The relationship of $\gamma$-ray nebulae to radio pulsars in
the H.E.S.S. Galactic Plane Survey (reproduced from \cite{493}). Left: 
number of radio pulsars where coincident $\gamma$-ray emission exists 
(thick histogram), as a function of the spin-down flux ($\dot{E}/d^{2}$)
in erg\,s$^{-1}$\,kpc$^{-2}$. Thin lines show the whole population, and 
an estimate of the number of chance associations expected. Right: 
the fraction of radio pulsars which appear to have associated 
TeV $\gamma$-ray PWN, again as a function of spin-down flux. See \cite{493} 
for details.
}
\label{fig:pwnpop}
\end{center}
\end{figure*}

PWN have now emerged as the largest population of identified TeV 
sources (see table~\ref{tab:gal}). As the number of extended 
VHE $\gamma$-ray sources along the Galactic Plane has increased
the likelihood of chance associations with pulsars is now
far from negligible. At this conference, the H.E.S.S. collaboration
presented a systematic search for coincidences between sources
detected in the H.E.S.S. galactic plane survey with radio pulsars \cite{493}.
As is evident from figure~\ref{fig:pwnpop} there is a clear excess
of $\gamma$-ray nebulae in positional coincidence with 
high spin-down luminosity pulsars (those with $\dot{E}/d^{2}$ above $\sim$10$^{35}$
erg\,s$^{-1}$\,kpc$^{-2}$) over the expectations for chance coincidences.
The implied efficiency in the conversion of spin-down power 
into TeV $\gamma$-ray production for these pulsars is around 1\%.

Six new $\gamma$-ray sources coincident with high spin-down luminosity pulsars
were presented here by the H.E.S.S. collaboration. These probable PWN can be
roughly categorised by the characteristic spin-down age 
$\tau_{\rm C}\equiv  P/2\dot{P}$
of the associated pulsar. Two of the associated pulsars are very young 
(i.e. similar to the Crab pulsar with $\tau_{\rm C} \sim$1000 years):
PSR\,J1846$-$0258 in Kes\,75 and G\,21.5$-$0.9 \cite{1057}. The remaining four 
have $\tau_{\rm C} \sim$10$^{4}$ years:
HESS\,J1718$-$385,\cite{494}
HESS\,J1809$-$193 \cite{1036}
HESS\,J1357$-$645 \cite{269}
and HESS\,J1912+102 \cite{269}, 
see \cite{1091} for a discussion of these objects.

Despite the large number of new PWN candidates, the most significant
recent discovery in this area is that of energy dependant morphology
in HESS\,J1825$-$137 \cite{389}.  The new data from the
H.E.S.S. collaboration show that the $\gamma$-ray emission `shrinks'
towards the pulsar PSR\,B1823$-$13 at high energies. Such behaviour
has been seen before in X-ray synchrotron emission and has been
interpreted as evidence for the cooling (energy-losses) of $>$ TeV
electrons. The discovery of this effect in $\gamma$-rays provides us
with a new tool with which to investigate the high energy particles in
these objects.

A final PWN candidate worthy of note here is the \emph{C3} `hot-spot' 
detected using MILAGRO at the position of the Geminga pulsar \cite{735}.
Whilst the signal is estimated at only 2.8$\sigma$ after correcting
for statistical trials  (and 5.1$\sigma$ pre-trials) the coincidence 
with a powerful EGRET pulsar is compelling. The MILAGRO source has an
apparent spatial extent of $\sim$2.8$^{\circ}\pm0.8^{\circ}$. Assuming
this object lies at the Geminga distance of $\sim$300~pc, its intrinsic
size is $\sim$15~pc, comparable to that of other more distant PWN such as 
MSH\,15-52. Unfortunately, a source of this angular size will be extremely 
difficult to verify with current air-Cherenkov telescopes due to their 
restricted FoV.

No ground-based instrument has so far provided convincing evidence for
pulsed $\gamma$-ray emission from a radio pulsar. The highest energy
pulsed photons are those detected using EGRET at $\sim$10~GeV. As
pulsed emission at higher energies is predicted in some scenarios,
several groups have pursued pulsed emission searches from prominant
GeV pulsars. Upper limits resulting from these searches were presented
by the PACT~\cite{517}, 
Tibet AS$\gamma$~\cite{844},
H.E.S.S.~\cite{572} and 
STACEE~\cite{830} collaborations. Tantalising
hints of a pulsed signal from the Crab were presented by the MAGIC
collaboration. A 2.9 $\sigma$ pulsed excess is seen in the phases of
peak $>$100~MeV emission \cite{1078}. More data is clearly required 
to confirm this potentially very important result.

\subsection{Binary Systems}

Much controversy surrounds the early claims of TeV (and indeed PeV)
emission from X-ray binary systems but recent progress has led to a
catalogue of three well established $\gamma$-ray binaries. The first
of these is PSR\,B1259$-$63 / SS\,2883 a 3.4~year period binary of a
pulsar in an eccentric orbit around a Be-star from which variable TeV
emission was detected during its periastron passage in early 2004. The
TeV emission from this object is thought to be associated with the
pulsar wind and its interaction with the radiation field and material
around the Be-star.  The 2nd periastron to be observed by TeV
instruments has just occurred (in July 2007) and will also be closely
observed by several X-ray satellites. The H.E.S.S. collaboration
presented a detection of this source in observations just befiore the
conference, and plans for upcoming multi-wavelength observations
\cite{553}.  The two remaining systems are both much closer binaries,
for which the mass and indeed the nature of the compact object are
unknown. The first of these to be discovered was LS\,5039, detected in
the H.E.S.S. galactic plane survey.  The emission of LS\,5039 is
clearly periodic and it has been possible to extract a binary period
of $3.9078 \pm 0.0015$ days (cf the optical period of $3.90603 \pm
0.00017$) from the $\gamma$-ray data alone \cite{1305}. Furthermore,
the $\gamma$-ray spectrum of the object clearly varies as a function
of phase, with a softening when the compact object lies behind its
companion that may be indicative of $\gamma$-$\gamma$ absorption or
cascading. The second well established TeV emitting X-ray binary
system is LS\,I\,+61\,303, discovered by the MAGIC collaboration in
2006. This object has been the subject of subsequent observing
campaigns with VERITAS \cite{700,351} 
and MAGIC \cite{574} which were
presented here. Whilst LS\,I\,+61\,303 is certainly variable, it is
not yet clear if it is strictly periodic, with good phase coverage
hampered by an orbital period (26.5~days) close to that of the lunar
cycle.

As the nature of the compact object is unknown in these two systems,
it is not clear if the emission is due to a relativistic outflow from
a neutron star (i.e. rotation powered as PSR\,B1259$-$63 / SS\,2883
seems to be) or accretion on to a black hole or a neutron star which
drives a relativistic jet. See \cite{Mirabel:Binaries} for a
discussion.  In this context, the recent evidence for TeV emission
from the binary Cyg\,X-1, which contains a $>$13$M_{\odot}$ black
hole, is very exciting: such a system must be powered by accretion
rather than rotational energy.  A $4.9\sigma$ excess is seen in one 79
minute period in the 40 hours of MAGIC observations \cite{551}. The
apparent TeV outburst occurred during a period of enhanced X-ray
activity, but there does not appear to be a correlation between X-rays
and $\gamma$-rays on short timescales. The estimated post-trials
significance of this signal is $4.1\sigma$, but as was discussed at
the conference, the assessment of statistical trials is not
straight-forward in this case. For this reason, the status of Cyg\,X-1
as a TeV emitter cannot yet be considered as proven beyond doubt
(hence its omission from table~\ref{tab:gal}).  A confirmation of this
signal using VERITAS or via further MAGIC observations is therefore
highly desirable.

\subsection{The Galactic Centre}
\label{sec:gc}

The central $\sim$100~pc of our galaxy is host to a wide range of
potential TeV emitting objects. The most exotic of these,
and also the most widely discussed, are the supermassive black hole
Sgr A$^{\star}$ and a hypothetic cusp of self-annihilating dark 
matter. TeV emission from close to Sgr A was discovered independently using the
Whipple \cite{GC:Whipple04}, 
CANGAROO \cite{GC:CANGAROO} and 
H.E.S.S. \cite{GC:HESS} instruments in 2004.
In addition to this point-like source (HESS\,J1745$-$290), 
diffuse emission correlated with the giant molecular clouds (GMCs) 
of the central region was discovered using H.E.S.S. in 2006 \cite{GC:hessDiffuse}.
The theoretical work on the $\gamma$-ray emission of the galactic
centre (GC) region at this conference was focused primarily on the
diffuse emission. Moskalenko et al discussed CRs injected
from the supernova remnant Sgr A East, propagating through and
radiating in the GMCs of the GC\cite{744}.
Erlykin and Wolfendale considered an origin of the emission as a consequence
of a succession of SNRs in the region over the past $10^{5}$ years \cite{11}.

Whilst no new experimental results on the diffuse component were presented,
there were four contributions 
on the central source HESS\,J1745$-$290. Over the past 2--3 
years there has been a major effort to drive down the 
systematic errors on pointing of the H.E.S.S. telescopes,
resulting in an extremely precise localisation of the TeV
emission at the GC \cite{286}, the reported centriod of the
emission has $6''$ statistical and $6''$ systematic errors.
The new position effectively excludes the SNR Sgr A East 
as the dominant source of the TeV emission. The
PWN candidate G\,359.95$-$0.04 and the supermassive black hole 
remain as the most likely candidates.

The observation of a major X-ray flare from Sgr~A$^{\star}$ during
simultaneous measurements with H.E.S.S. and Chandra in July 2005 \cite{463}
provides a unique opportunity to test the association of the TeV
source with the supermassive black hole. There was no evidence for
an increase in the $\gamma$-ray flux during this event, constraining 
any flaring TeV component to be less than 100\% of the steady component
during the $\approx$30 minutes of the flare. A search of the full H.E.S.S. data
set yielded only upper limits on variability and QPOs \cite{1021}.
These results limit models for HESS\,J1745$-$290 as arising from
acceleration at Sgr A$^{\star}$ to those in which the accelerated particles
propagate rather far ($\sim$1 pc) from the supermassive black hole before
losing significant energy (see for example \cite{aharonianNeronov}).
Limits on a dark matter annihilation component to the spectrum
of HESS\,J1745$-$290 were also presented \cite{886}.

\subsection{Unidentified Sources}

The majority of galactic TeV $\gamma$-ray sources have no clear
counterpart at other wavelengths. This situation likely results from a
combination of experimental and physical considerations. A primary
reason is certainly that many of these sources are widely extended and
may have morphology that differs significantly from that at other
wavelengths. There are two basic categories of unidentified TeV
source: 1) sources where there is a candidate for the emission, but no
strong evidence to support an association (for example in several
cases there is an ambiguity between SNR shell emission and PWN
emission due to a lack of angular resolution and/or statistics) and 2)
sources where \emph{no} good candidate exists at sub-$\gamma$-ray
wavelengths (TeV sources with GeV associations cannot be considered as
identified) which have sometimes been referred to as `dark
sources'. The first example of the latter type was TeV\,J2032+4130,
discovered by the HEGRA collaboration in 2002 and has now been
confirmed using MAGIC \cite{532}. The second such object was
HESS\,J1303$-$631 , serendipitously discovered using H.E.S.S. in 2004 and
recently confirmed using CANGAROO-III~\cite{320}.  Many more objects in
this class have followed.  A summary of sources with no good
counterpart at any wavelength below the $\gamma$-ray was presented
here by the H.E.S.S. collaboration \cite{400}, including six TeV
sources newly discovered in the H.E.S.S. galactic plane survey.
A further unidentified H.E.S.S. source: HESS\,J0632+057, was 
recently discovered close to the Monoceros Loop SNR and is unusual
in its point-like nature \cite{581}.

\begin{figure}
\begin{center}
\includegraphics[width=0.5\textwidth]{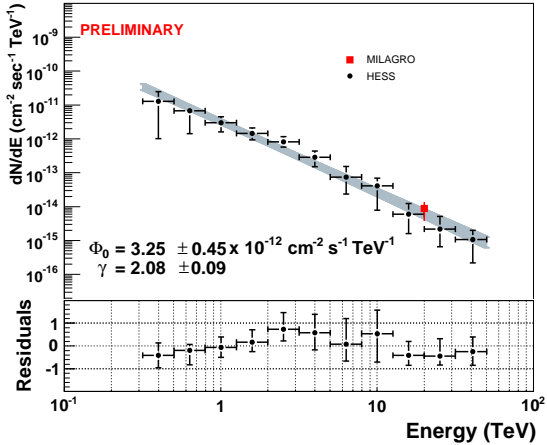}
\caption{Energy spectrum for MGRO\,J1908+06 from MILAGRO and
H.E.S.S. data. The lower panel shows residuals to a power-law fit. 
Reproduced from \cite{1316}
}
\label{fig:1908}
\end{center}
\end{figure}

Very recently the MILAGRO collaboration has added three more objects to
this list: MGRO\,J2031+41, MGRO\,J2019+37 and MGRO\,J1908+06
\cite{735}.  These objects have fluxes approaching that of the Crab
Nebula above 20~TeV and one (MGRO\,J2031+41) is significantly extended
beyond the $\sim$1$^{\circ}$ angular resolution of MILAGRO.  Flux upper
limits on \emph{point-like} emission from MGRO\,J2019+37 from the
VERITAS \cite{1183} and 
MAGIC \cite{484} collaborations were presented
which exclude extrapolation of the MILAGRO fluxes down to $\sim$1~TeV
with typical $E^{-2.3}$ type spectra. However, as this source is
probably extended (best fit diameter $1.1\pm0.5^{\circ}$) these point
source ($<$0.1$^{\circ}$) limits may not be meaningful. Indeed,
MGRO\,J2019+37 has now been confirmed using Tibet AS$\gamma$
\cite{548}, an instrument with comparable resolution to MILAGRO. The
detection of MGRO\,J1908+06 by the H.E.S.S. collaboration presented
here \cite{1316} is the first confirmation of a source detected by a
non-IACT instrument by an IACT system. The excellent agreement on the
$\sim$20~TeV flux of this source, illustrated in
figure~\ref{fig:1908}, provides further confidence in the MILAGRO
detections.  Figure~\ref{fig:1908} also illustrates the power of the
imaging technique for spectral measurements. The H.E.S.S. data shown
were obtained in just a few hours, in comparison to the 7 years
integration of the single MILAGRO point on this $\sim$0.5$^{\circ}$
diameter source.  Nevertheless, wide field of view instruments such as
MILAGRO are certainly complementary to the existing narrow FoV IACTs
for the detection of extended emission and such high duty cycle
instruments have a clear advantage in the search for transient
phenomena.

One exotic explanation that has been put forward for these
unidentified sources is that they originate in the annihilation of
dark matter in localised `clumps'.  A major difficulty of this
explanation is the energy spectrum of the sources, for example HESS\,J1303$-$631
\cite{890}. The more prosaic explanation put forward is that they 
originate in the collisions of cosmic-ray hadrons, with little
emission at other wavelengths (in contrast to electrons which
typically produce comparable fluxes in synchrotron emission). However,
in this scenario the acceleration site for these CRs remains a
mystery.

Perhaps the most significant of the new sources without a clear
counterpart is HESS\,J1023$-$575 \cite{211}. This object is coincident
with the massive stellar cluster Westerlund\,2 the second most massive
young cluster in our galaxy.  Whilst this association may be
coincidental, the colliding winds of stars in this cluster can
certainly provide the energy required to produce the $\gamma$-ray
emission and acceleration in such objects seems plausible (see for
example \cite{Torres:Winds} and references therein). As such,
HESS\,J1023$-$575 may well be the first of a new class of galactic
$\gamma$-ray sources. As well as the conventional $\gamma$-ray
production mechanisms discussed above, it was suggested at the
conference that the photo-disintegration of nuclei may play an
important role in this object and in other high radiation-field
environments \cite{407}.

\section{OG 2.3: Extragalactic Sources}

\subsection{AGN}

Active galactic nuclei (AGN) are thought to harbour 
actively accreting supermassive black holes which drive
relativistic jets into their environments. 
The \emph{blazar} subclass of AGN is characterised by rapid variability and
high energy ($>$0.1~GeV) emission. These objects are thought to
represent AGN with jets aligned very closely ($<$10$^{\circ}$) with 
the line of sight to the observer, resulting in greatly enhanced fluxes
through beaming effects. Blazars were the dominant
source class detected with EGRET at GeV energies and beginning in
1992 with Mrk 421, a class of higher energy peaked \emph{TeV blazars}
has been established. The spectral energy distribution of blazars
is double peaked with a minimum  typically somewhere in the hard X-ray 
to soft $\gamma$-ray energies ($\sim$1~MeV). The most common explanation for
the two components is as synchrotron and inverse Compton radiation of
a population of energetic electrons within a region with bulk relativistic
motion along the jet. The high energy component has also been interpreted
as due to accelerated hadrons (via several different radiation processes). 
These explanations are of particular relevance to the cosmic ray field
as AGN are one of the primary candidates for the acceleration of the 
ultra-high energy cosmic rays (those with $E>10^{19}$ eV).

The theoretical work on AGN at this conference included studies
of time variability in inverse Compton $\gamma$-ray spectra \cite{304} and on
the effects of jet expansion on blazar emission properties \cite{682}. The 
vast majority of contributions were, however, experimental in nature.
There were two main experimental highlights: the discovery of seven new
TeV blazars and the measurement of extreme $\gamma$-ray variability 
in three previously known objects.

Table~\ref{tab:agn} summarises the known TeV AGN, including the seven new
objects presented at this conference. There are now sufficient numbers
of these objects to allow population studies, a project which is now
underway. At this conference a study was presented exploring the relationship
of the TeV emission to the properites of the active galaxy, including the black hole
mass~\cite{70}. The sources of table~\ref{tab:agn} are ordered by redshift,
illustrating the recent progress made in measurements of more distant objects.
Aside from their interest as particle accelerators, the TeV blazars
are important beyond the field of high energy astrophysics as they
have been used place constraints on the star-formation history of
the universe. The energy-dependent absorption of $\gamma$-rays via 
pair-production on the extragalactic background light (EBL) can be used
to derive limits on the energy density of this photon field and hence
on the integrated radiation history of galaxies. Conversely this 
absorption places an energy dependent horizon on $\gamma$-ray observations.
An optical depth of $\tau=1$ is reached at a redshift of $\sim$0.1 for
1~TeV $\gamma$-rays. Only relatively recently have experiments with substantial
sensitivity in the 0.05--1 TeV range existed, leading to a rapid expansion in the
number of $z>0.1$ TeV blazars. 

Three of the these new objects were discovered using the
H.E.S.S. instrument: PKS\,0548-322, 1ES\,0229+200 and
1ES\,0347$-$121. These objects are all classified as high energy
peaked BL Lac objects or HBLs, based on the position of the peak in
the synchrotron spectrum. The relatively hard energy spectra measured
for 1ES\,0229+200 and 1ES\,0347$-$121 (photon indices $\sim$2.5 and
$\sim$3.1 respectively) make them particularly useful for constraining
the EBL.  Under the assumption that the intrinsic spectrum of these
objects has a photon index not less than 1.5 (that expected for
inverse Compton radiation of an $E^{-2}$ electron spectrum radiating
in the Thompson limit), limits on the mid- and near Infra Red were
presented that approach the lower limits from galaxy counts at these
wavelengths \cite{568}.  Combined EBL limits using all previously
known TeV blazars were also presented here \cite{1045}.

\begin{table*}
\begin{center}  
\begin{tabular}{|l|l|l|l|l|l|} \hline  
Object       &  Discovered &  Year & $z$   & Class   & Contrib. \\\hline 
M\,87         &  HEGRA      &  2003 & 0.004 & LINER  &  \cite{499,1120,756}   \\
Mrk\,421      &  Whipple    &  1992 & 0.031 & HBL    &  \cite{665,748,234}   \\
Mrk\,501      &  Whipple    &  1996 & 0.034 & HBL    &  \cite{1098,234,170}   \\
1ES\,2344+514 &  Whipple    &  1998 & 0.044 & HBL    &  \cite{71}   \\
Mrk\,180      &  MAGIC      &  2006 & 0.046 & HBL    & \cite{936}    \\
1ES\,1959+650 &  TA         &  2002 & 0.047 & HBL    & \cite{911}    \\
BL\,Lac       &  MAGIC      &  2006 & 0.069 & LBL    & \cite{946}    \\
PKS\,0548$-$322 &  HESS       &  2006 & 0.069 & HBL    &  \cite{331}   \\
PKS\,2005$-$489 &  HESS       &  2005 & 0.071 & HBL    &  \cite{511}   \\
PKS\,2155$-$304 &  Durham      &  1999 & 0.116 & HBL    &  \cite{714,1221,205,243} \\ 
H\,1426+428   &  Whipple    &  2002 & 0.129 & HBL    &  \cite{779}   \\
1ES\,0229+200 &  HESS       &  2007 & 0.140 & HBL    &   \cite{568}   \\
H\,2356$-$309   &  HESS       &  2005 & 0.165 & HBL    &  \cite{511}   \\
1ES\,1218+304 &  MAGIC      &  2005 & 0.182 & HBL    &  \cite{564,304}   \\
1ES\,1101$-$232 &  HESS       &  2005 & 0.186 & HBL    &  \cite{555}   \\
1ES\,0347$-$121 &  HESS       &  2007 & 0.188 & HBL    &  \cite{568}   \\
1ES\,1011+496 &  MAGIC      &  2007 & 0.212 & HBL    &  \cite{936}   \\
PG\,1553+113  &  HESS       &  2005 & $>$0.25 & HBL  &  \cite{1210,72}   \\
3C\,279       &  MAGIC      &  2007 & 0.536 & FSRQ   &  \cite{1019}   \\
\hline
\end{tabular}
\end{center}
\caption{The known very high energy $\gamma$-ray emitting AGN.
The instrument used for the first VHE detection is given together 
with the year of discovery, the redshift and the object class.
The main contributions to this conference
containing VHE data are listed for each object.
}
\label{tab:agn}
\end{table*}

\begin{figure*}
\begin{center}
\includegraphics[width=0.99\textwidth]{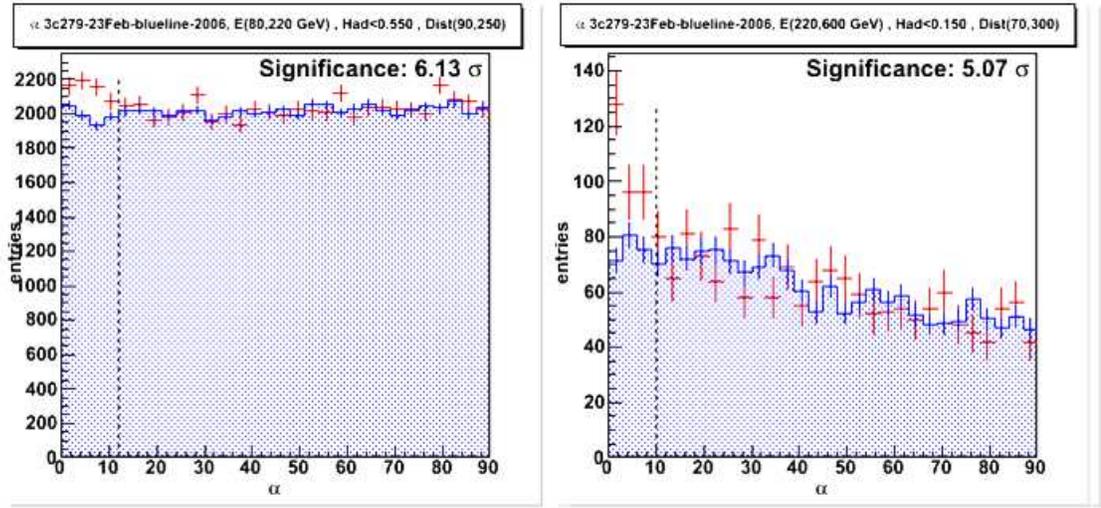}
\caption{VHE emission from 3C\,279 on the 23$^{\rm rd}$ of February 2006.
\emph{Alpha-plots} for \emph{on-} (points) and \emph{off-} (histograms) 
data collected using the MAGIC telescope in two energy bands:
left: 90-220 GeV and right: 220-600 GeV. 
$\alpha$ is the angular distance between the major axis of 
a Cherenkov image seen in the camera and the line connecting
the image centriod to the position of the target source.
}
\label{fig:3c279}
\end{center}
\end{figure*}

The four new objects presented by the MAGIC collaboration are all
interesting for three rather different reasons. Firstly, the detection
of BL Lacertae is important is this is the first low energy peaked BL
Lac object (LBL) to be detected using a ground-based instrument
\cite{946}. It seems likely that a large number of such sources may be
detected by lower threshold instruments such as HESS-II and MAGIC-II.
The MAGIC discoveries of VHE emission from both Mrk 180 and
1ES\,1011+496 arose from observations triggered by optical activity
\cite{936}.  The implied optical/TeV connection may be important not
just for our understanding of these objects but on the practical
grounds that optical monitoring of a large sample of AGN is much
easier to achieve than a X-ray campaign on a similar
scale. 1ES\,1011+496 ($z=0.212$) was also (briefly) the most distant
known TeV source with a well established redshift, displaced by the
MAGIC discovery of $\gamma$-ray emission from 3C\,279 \cite{1019}
first announced at this conference.  The discovery of TeV emission
from the GeV bright blazar 3C\,279 is important in two respects:
firstly as it marks a major step forward in redshift for ground-based
instruments (to $z=0.536$) and secondly as this object belongs to a
rather different class of AGN: the Flat Spectrum Radio Quasars.
3C\,279 was the brightest extragalactic object detected using EGRET
and is hence certainly a GeV rather than a TeV blazar. With the
upcoming launch of GLAST, 3C\,279 may become the only object for which
simultaneous GeV and TeV measurements are possible on $\sim$1~hour
timescales. The MAGIC signal from 3C\,279 (shown in
figure~\ref{fig:3c279}) consists of one night of significant emission
from a ten night observation. The signal is at the $6.1\sigma$ level
(without accounting for statistical trials) in an energy band from
80--220 GeV, and at the $5.1\sigma$ above 220 GeV. The signal in the
higher energy band is particularly surprising given the redshift of
this object. The energy spectrum of this source will be extremely
interesting from the perspective of EBL absorption. Given the
importance of this detection, caution is necessary and a very careful
assessment of statistical trials (notoriously difficult for variable
sources) and systematic effects is clearly needed. However, given the
strength of the signal, and its independent confirmation in a second
energy band, it seems highly likely that 3C\,279 is a VHE $\gamma$-ray
source. As 3C\,279 is readily accessible from both hemispheres a
confirmation should be possible rather quickly and this object should
be a prime candidate for coordinated monitoring with MAGIC, VERITAS
and H.E.S.S.

\begin{figure}
\begin{center}
\includegraphics[width=0.5\textwidth]{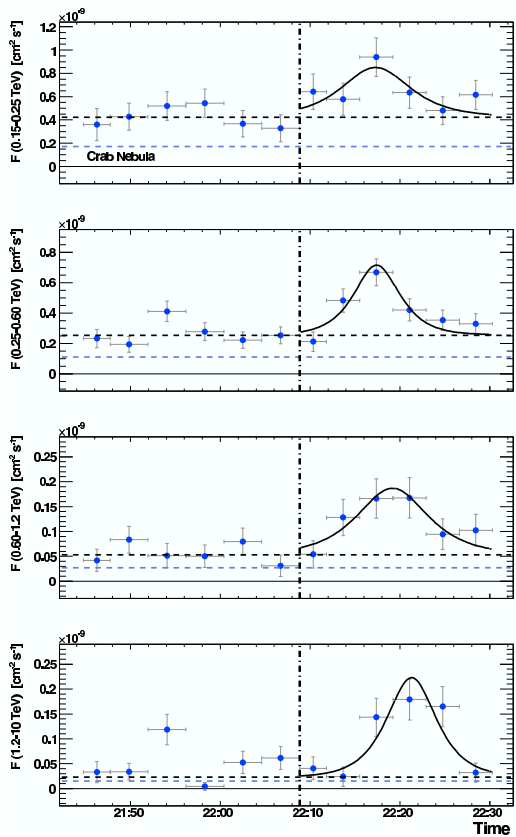}
\caption{MAGIC light-curves of a flare from Mrk\,501
on the 9$^{\rm th}$ of July 2005. The data are subdivided in to 4
energy bands. The flux of the Crab Nebula in each band is indicated by
a dashed horizontal line.  Reproduced from \cite{1098}.  }
\label{fig:magic_lag}
\end{center}
\end{figure}

Since the 29$^{\rm th}$ ICRC spectacular flaring activity has been
seen in two TeV blazars: Mrk\,501 \cite{1098} and PKS\,2155$-$304
\cite{1221}.  The Mrk\,501 activity observed using MAGIC in July 2005
was the first major outburst observed by a instrument of the more
sensitive new generation. As such the temporal and spectral resolution
possible surpassed that of previous measurements. The highlight of
these observations is the detection of very fast ($\sim$2~minute flux
doubling time) variability with a significant lag between photons of
different energies (see figure~\ref{fig:magic_lag}). Such lags are a
potentially powerful diagnostic of acceleration and energy loss
processes and the short timescales involved place tight limits on the
size of the emitting region and the Doppler factor of the jet ($\delta
> 16$ is inferred from these measurements
\cite{1098}). 

\begin{figure*}
\begin{center}
\includegraphics[width=0.99\textwidth]{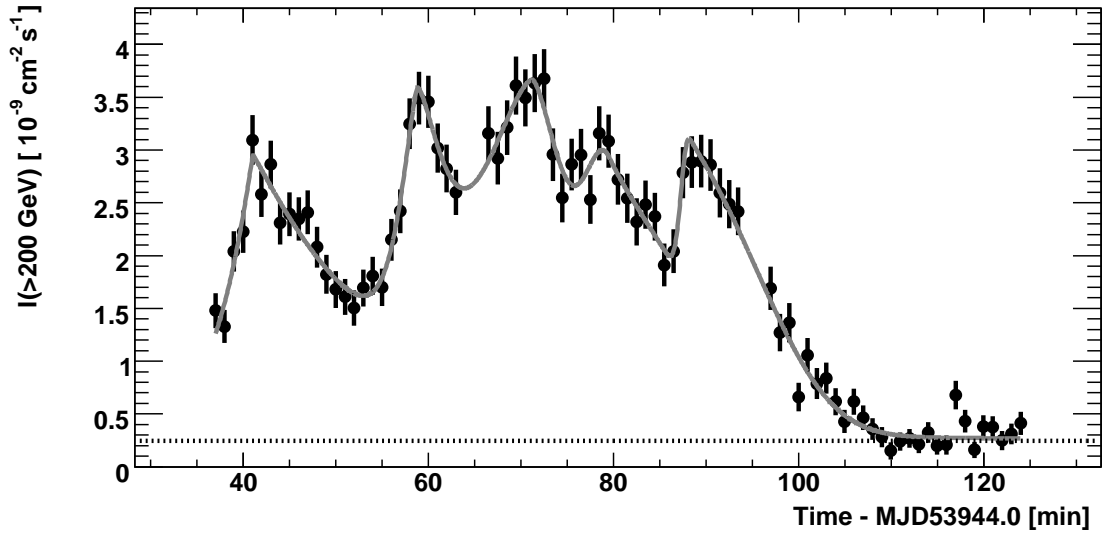}
\caption{H.E.S.S. VHE flux light-curve from a flare of 
PKS\,2155$-$304 on the 28$^{\rm th}$ of July 2006.  The one-minute
binned data are fit to a multi-component Gaussian (smooth curve).  The
flux of the Crab Nebula is indicated as a dashed line for
comparison. Reproduced from \cite{1221}.  }
\label{fig:2155flare}
\end{center}
\end{figure*}

The activity of PKS\,2155$-$304 observed using H.E.S.S. in July 2006
was even more dramatic \cite{1221}. Figure~\ref{fig:2155flare} shows
the light curve of the night with the highest flux, in which the
emission reached fluxes more than two orders of magnitude higher than
the quiescent flux of this object. Short timescale variability is
clearly evident in figure~\ref{fig:2155flare} and the best measured
individual flare is the first of the night with a best fit rise-time
of $173\pm23$ seconds. No evidence for energy dependent time-lags was
presented at the conference.

A possible `spin-off' of these measurements of fast variability in
distant objects is to constrain any energy-dependence of the speed of
light and hence probe the energy scale of Quantum Gravity effects. See
\cite{592} for details.

\begin{figure}
\begin{center}
\includegraphics[width=0.49\textwidth]{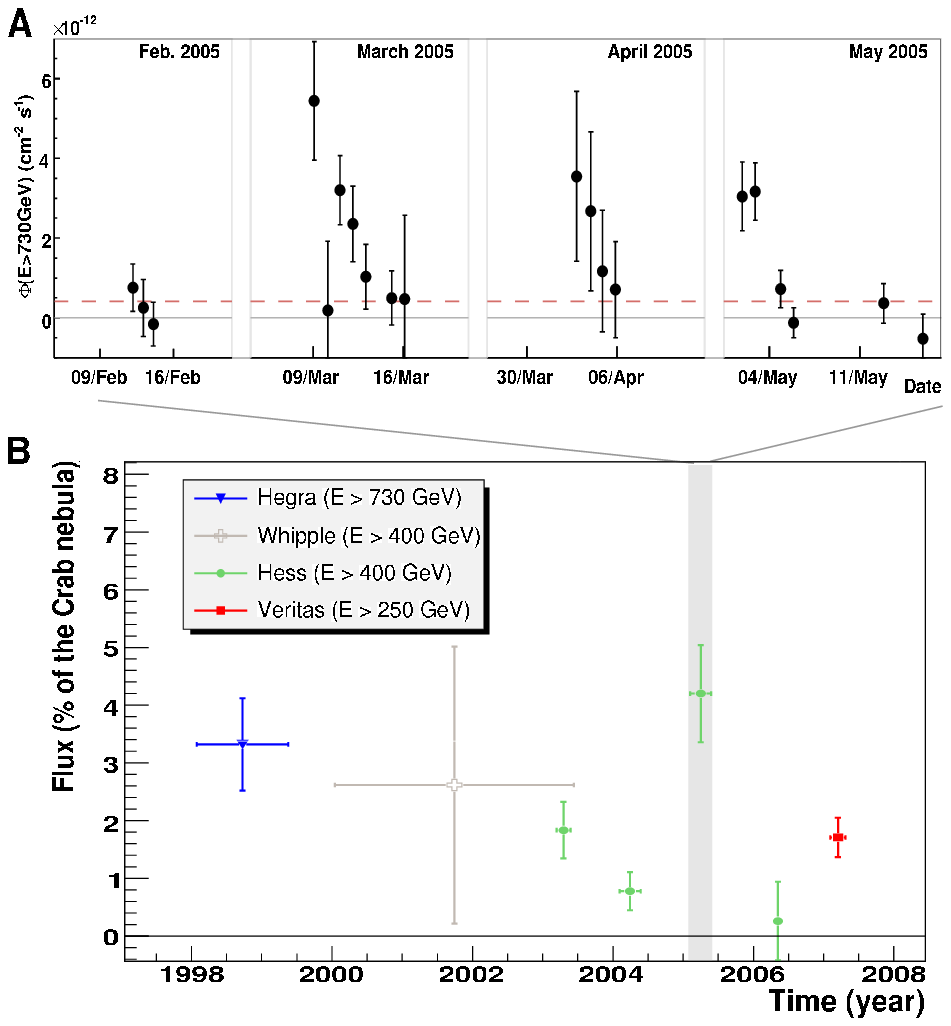}
\caption{Long and short-term variability in the TeV emission of M\,87.
A) Short-term variability seen in the light-curve of M\,87 using
H.E.S.S. in 2005, reproduced from \cite{499} and B) Long-term
variability as seen using HEGRA, Whipple, H.E.S.S. and VERITAS,
reproduced from \cite{756}.  }
\label{fig:m87}
\end{center}
\end{figure}

The only non-blazar known to emit TeV photons is the 
nearby ($z=0.004$) radio galaxy M\,87,
the core of which harbours the most massive known black hole in
the nearby universe. The angle between the line-of-sight and the
jet axis appears to be $\sim$30$^{\circ}$ in this system, in 
contrast to the $<$10$^{\circ}$ inclination angles of the blazars.
Given the reduced beaming effects in such a system and the mass
of the black hole, the two day timescale variability discovered 
using H.E.S.S. \cite{499} is particularly surprising. Causality
arguments have been used to derive a limit of $5 \delta R_{s}$ on
the size of the emission region, where  $\delta$ is the Doppler factor
of the source and $R_{s}$ is the Schwarzschild radius of the 
supermassive black hole. Figure~\ref{fig:m87} shows the light-curve
of M\,87 on long (year) and short (day) timescales including data
from several VHE instruments. The most recent data shown are the
$5.1 \sigma$ detection of this source using VERITAS earlier 
this year \cite{756}.


\subsection{Potential Extragalactic TeV Source Classes}

Although AGN are the only extragalactic TeV source class identified so
far, there are several other object classes with TeV fluxes which may
be reachable with current or near future instruments. The primary
target class in terms of investment of observing time seems to be
Starburst Galaxies and their cousins the ultra-luminous infra-red
galaxies or ULIRGs.  These objects present the possibility of
exploring CR acceleration associated with stellar life-cycles
(normally assumed to occur in SNRs) in an integrated fashion. Upper
limits on the Starbursts in nearby galaxies NGC\,253 and M\,83 were
presented by the H.E.S.S. collaboration \cite{468}, and on the ULIRG
Arp\,220 using MAGIC \cite{1288}. These limits are already deep enough
to challenge the simplest scenarios for cosmic acceleration and
propagation in these objects and further observations remain well
motivated.

As the largest gravitationally bound structures in the universe,
Galaxy clusters are of crucial importance in many areas of
astrophysics and cosmology. As the escape and energy-loss timescales
of ultra-relativistic hadrons in these systems is longer than a Hubble
time \cite{BreitschwardClusters} $\gamma$-ray observations of clusters
could potentially probe the integrated CR acceleration history of
these objects. Possible sites of injection of CRs into the
intra-cluster medium include shocks associated with large scale
structure formation (merger and/or accretion shocks), stellar
processes within cluster member galaxies (e.g. SNR) and AGN outbursts
\cite{882}. Flux upper limits from the H.E.S.S. and CANGAROO-III
instruments were presented on the galaxy clusters Abell 496 and Coma
\cite{535} and on Abell 4038 and 
Abell 3667 \cite{829}.

The increasingly deep upper limits on the most prominent members of
these source classes suggest that non-beamed emission from
extragalactic sources may be difficult for current TeV instruments to
detect. Nevertheless, it seems likely that these source classes lie
within the reach of near future instruments such as GLAST and the
second phase instruments of H.E.S.S. and MAGIC, and could have a huge
impact on the cosmic ray field.

Other extragalactic objects considered include globular clusters \cite{26}
and possible dark matter annihilation in dwarf galaxies \cite{1076}.

\section{OG 2.4: Gamma-Ray Bursts}

Gamma-ray bursts (GRBS) are widely understood as originating in
relativistic `fireballs' following the core-collapse 
of massive stars and/or the coalescence of two compact objects.
A high energy component (possible from inverse Compton scattering
of high energy electrons) may exist in these bursts and
emission up to $\sim$ 20 GeV was seen using EGRET,
but as of yet no completely convincing case for TeV emission 
from a GRB exists. Some theoretical work on GRBs was presented at 
this conference \cite{1134,1079,1141,1168} but the majority of 
the contributions were experimental in nature and most of these 
presented fluence limits on individual GRBs in the TeV energy range.

The Gamma-ray bursts Coordinates Network (GCN) provides automatic
alerts to subscribing ground-based instruments following the detection
of a GRB by a satellite based detector. Currently, most such alerts
are triggered by the Swift satellite but HETE-2 and Integral also
provide alerts. Most TeV instruments subscribe to this system and
respond to alerts where possible. The response time of Cherenkov
telescopes is limited in principle only by the typical GCN delay of
a few seconds plus the slewing time of the telescope(s). The MAGIC
telescope was designed in a light-weight manor with the specific aim
of slewing rapidly to GRBs and has a speed of $\sim$5$^{\circ}$/s. 
The more heavily built H.E.S.S. and VERITAS telescopes slew at
$\sim$2$^{\circ}$/s and $\sim$1$^{\circ}$/s respectively. Four pointing
instruments presented upper limits from their GRB programs:
MAGIC \cite{566}, 
H.E.S.S. \cite{466}, 
VERITAS \cite{406} and
STACEE \cite{409}. H.E.S.S. is unique in having observed a burst
with zero delay: GRB\,060602B occurred serendipitously at  
2.5$^{\circ}$ from the pointing direction of the array \cite{464}.
However, this burst has may in fact have been an X-ray flash 
of galactic origin. After this, the fastest response of a pointed
instrument to a GRB is the MAGIC of  GRB 050713a, starting 
40~seconds after the burst trigger, but in the absence of a redshift
measurement the fluence upper limit obtained is hard to interpret.

Very wide field instruments such as MILAGRO and Tibet AS$\gamma$
have clear advantages in the search for TeV emission from GRBs.
Their close to 100\% duty cycle and very large field of view
ensure that prompt VHE emission from many bursts can be tested.
The disadvantage of somewhat poorer fluence sensitivity for this
instruments is probably outweighed by the advantage of a zero 
response time, but this obviously depends on the (unknown) 
time profile of the high energy component of the burst.
The MILAGRO collaboration presented upper limits from two 
approaches probing different energy bands \cite{1176,689}.

All VHE instruments face a severe difficulty in the limited
redshift range to which they are sensitive due to EBL absorption.
Only a small fraction of GRBs occur at small enough distances
and only a fraction of these will have measured redshifts.
It may therefore require considerable patience to measure 
$>100$ GeV emission from GRBs even if this component exists.
An instrument such as HAWC, with the advantages of MILAGRO,
but with a lower energy threshold providing much greater redshift
coverage, could be well suited to such studies \cite{1239}

\section{Summary}

\begin{figure*}
\begin{center}
\includegraphics[width=0.99\textwidth]{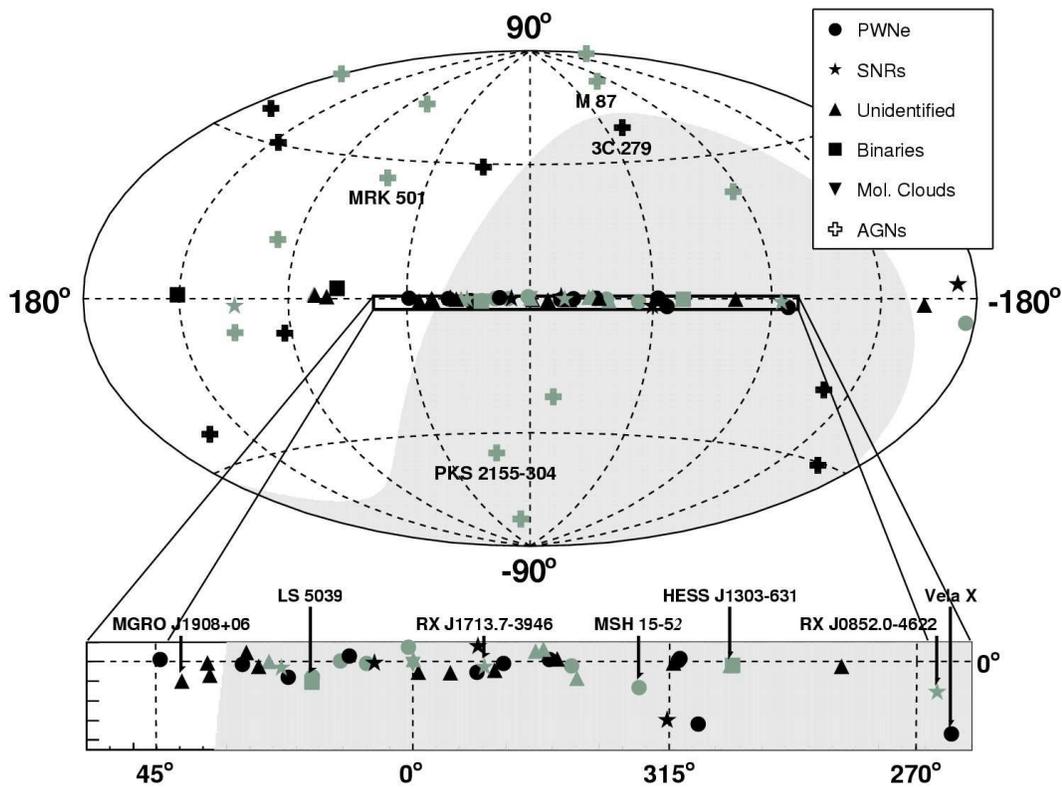}
\caption{The catalogue of known TeV sources as of the 30$^{\rm th}$ ICRC.
  The positions of known TeV emitters are shown in galactic
  coordinates.  Darker symbols indicate discoveries since the 29$^{\rm
  th}$ ICRC in 2005.  The shaded region indicates the part of the sky
  more readily accessible from the southern hemisphere (Declination
  $>$0$^{\circ}$), adapted from \cite{kappes}.  }
\label{fig:skymap}
\end{center}
\end{figure*}

It is clear that $\gamma$-ray astronomy is making rapid progress 
towards answering some of the important questions in cosmic ray
physics and contributing to several topics well outside the cosmic
ray field. It is already clear that GLAST will, if successfully 
deployed, have an enormous impact on the field, and it is highly 
likely that these results will dominate the next ICRC. 
For the moment the highlights are the results at $\sim$TeV energies. 
Figure~\ref{fig:skymap} shows the catalogue of known VHE $\gamma$-ray
sources as of mid-2007. The number of sources is very likely to 
grow from the current $\approx$71 to cross the 100 source threshold 
before the next ICRC. More importantly the number of established source 
\emph{classes} has grown, and there are hints of new source types
which may be established rather soon. The precision with which the 
brightest sources are being measured, for example all $6''$ errors 
on the centroid of the emission from the Galactic Centre, and 
the resolved energy dependent morphology in HESS\,1825$-$137, are perhaps
the best illustration of the progress made in the field.
Also extremely important is the detection of 3C\,279 using MAGIC,
marking a dramatic increase in the volume of the
universe accessible to ground-based $\gamma$-ray detectors.
With the completion of a major new VHE instrument, VERITAS, and
the ongoing construction of H.E.S.S.-II and MAGIC-II, it is likely
that this rapid progress will continue for some time to come.

\section*{Acknowledgements}
I would like to thank the ICRC organisers for inviting me to make this
summary and for an excellently organised conference. Many thanks to 
all the paper authors who provided material. I would also like to 
thank Richard White for his help with one of the figures and
Stefan Funk for his carefully reading of the draft.

\end{document}